\documentclass{aa}
\idline{1}{1}

\usepackage{graphicx}
\usepackage {amssymb}
\usepackage {mathrsfs}
\usepackage {txfonts}
\def\Teff  {$T_\mathrm{eff}$}
\def\S1  {S\,{\small I}}
\def\logg  {$\log g$}
\def\vt  {$v_t$}

\begin{document}

\title {First stars XIV. Sulfur abundances in extremely metal-poor stars
\thanks{Based on observations obtained with the ESO Very Large
Telescope at Paranal Observatory, Chile (Large Programme "First
Stars", ID 165,N-0276, P.I.: R. Cayrel.}
}

\author {
M.~Spite\inst{1}\and
E.~Caffau\inst{2,1}\and
S.M.~Andrievsky\inst{1,3}\and
S.A.~Korotin \inst{3}\and
E.~Depagne  \inst{4}\and
F.~Spite\inst{1}\and
P.~Bonifacio\inst{1}\and
H.-G.~Ludwig\inst{2}\and
R.~Cayrel\inst{1}\and
P.~Fran\c cois\inst{1}\and
V.~Hill\inst{5}\and
B.~Plez\inst {6} \and
J.~Andersen\inst {7,8} \and
B.~Barbuy\inst {9} \and
T.~C.~Beers\inst {10} \and
P.~Molaro\inst {11} \and
B.~Nordstr\"om \inst {7} \and
F.~Primas \inst {12} 
}

\institute{
GEPI Observatoire de Paris, CNRS, Universit\'e Paris Diderot, F-92195
Meudon Cedex France e-mail : {\tt monique.spite@obspm.fr}
\and  
Zentrum f\"ur Astronomie der Universit\"at Heidelberg,
Landessternwarte, K\"onigstuhl 12, 69117 Heidelberg, Germany
\and  
Department of Astronomy and Astronomical Observatory, Odessa National
University, and Isaac Newton Institute of Chile, Odessa branch, Shevchenko
Park, 65014, Odessa, Ukraine
\and  
Las Cumbres Observatory, Goleta, CA 93117, USA 
\and  
Observatoire de la C\^ote d'Azur, CNRS UMR6202, BP4229, 06304 Nice 
Cedex 4, France
\and  
GRAAL, Universit\'e de Montpellier II, F-34095 
Montpellier Cedex 05, France
\and  
The Niels Bohr Institute, Astronomy, Juliane Maries Vej 30,
DK-2100 Copenhagen, Denmark
\and  
Nordic Optical Telescope, Apartado 474, E-38700 Santa Cruz de 
La Palma, Spain
\and  
Universidade de S\~ao Paulo, Departamento de Astronomia,
Rua do Mat\~ao 1226, 05508-900 S\~ao Paulo, Brazil
\and  
Department of Physics \& Astronomy and JINA: Joint Institute for Nuclear
Astrophysics, Michigan State University, East Lansing, MI 48824, USA     
\and  
Istituto Nazionale di Astrofisica, Osservatorio Astronomico di
Trieste, Via Tiepolo 11, I-34143 Trieste, Italy
\and  
European Southern Observatory (ESO),
Karl-Schwarschild-Str. 2, D-85748 Garching b. M\"unchen, 
Germany
}

\date { }
\titlerunning{Sulfur abundance in extremely metal-poor stars.}
\authorrunning{M. Spite et al.}

\abstract
{Precise S abundances are important in the study of the early chemical evolution of
the Galaxy. In particular the site of the formation remains uncertain because, 
at low metallicity,  the trend of this $\alpha$-element versus [Fe/H]
remains unclear. Moreover, although sulfur is not bound significantly
in dust grains in the ISM, it seems to behave differently in
DLAs and old metal-poor stars.}
{We attempt a precise measurement of the S abundance in a
sample of extremely metal-poor stars observed with the ESO VLT
equipped with UVES, taking into account NLTE and 3D effects.}
{The NLTE profiles of the lines of multiplet 1 of S\,{\scriptsize I} were
computed with a version of the program MULTI, including opacity
sources from ATLAS9 and based on a new model atom for S. These profiles were fitted to the observed spectra.}
{We find that sulfur in EMP stars behaves like the other
$\alpha$-elements, with [S/Fe] remaining approximately constant below
[Fe/H]=--3.  However, [S/Mg] seems to decrease slightly with increasing
 [Mg/H]. The overall abundance patterns of O, Na, Mg, Al, S, and K
are most closely matched by the SN model yields by Heger \& Woosley.  The
[S/Zn] ratio in EMP stars is solar, as also found in DLAs.  We derive
an upper limit to the sulfur abundance [S/Fe] $< +0.5$ for the
ultra metal-poor star CS 22949-037.  This, along with a previously
reported measurement of zinc, argues against the conjecture that the
light-element abundance pattern of this star (and by analogy, the
hyper iron-poor stars HE 0107-5240 and HE 1327-2326) would be due to dust
depletion.}
{}
\maketitle
\section{Introduction} 

The determination of the abundance of sulfur in the atmosphere of old
Galactic stars is of interest for two primary reasons:

\noindent -First, since unlike most other heavy elements, sulfur is
generally considered not to be significantly bound in interstellar
dust grains, it is a good indicator of the chemical composition of the
ISM and thus allows a direct comparison between the abundances in the
DLAs (damped Ly$\alpha$ clouds) and early Galactic matter.

\noindent -Secondly, the formation of sulfur remains controversial.
Sulfur is generally considered as an $\alpha$-capture element such as
magnesium and calcium, formed preferentially in massive type II
supernovae, including those of the first generations of stars.  In
analogy with other $\alpha$-elements, [S/Fe] is expected to be
constant and positive at low metallicity (see e.g. Fran\c cois,
\cite{Fra87}, \cite{Fra88}).  However, the trend of [S/Fe] versus [Fe/H] in
the early Galaxy is still debated: Israelian \& Rebolo (\cite{IR01})
and Takada-Hidai et al.  (\cite{TTS02}) found [S/Fe] to rise with
decreasing [Fe/H], while Ryde \& Lambert (\cite{RL04}) and Nissen et
al.  (\cite{NCA04}, \cite{NAA07}) found [S/Fe] to remain flat.
Finally, Caffau et al.  (\cite{CBF05}, \cite{CSF10}), combining their
own data with a large sample from the literature covering the multiplets
1, 3, 6, and 8 of \S1  (see Table \ref{mults}), suggest a
bimodal distribution of [S/Fe] for $\rm[Fe/H]<-1.0$ to explain the
high value of [S/Fe] found in several metal-poor stars.

\begin {table}[ht] 
\caption {Parameters of the sulfur lines}  
\label {mults} 
\begin{center}
\begin{tabular}{c r r r r }
Mult. & wavelength  &  Transition                                & log $gf$  & $\chi_{ex}$\\
     & (nm) air    &                                            &           & (eV)       \\
\hline
1    & 921.2863    &4s-4p~~~$\rm ^{5}S\,^{0}_{2} - ^{5}P_{3}$      &  0.42     & 6.525      \\      
1    & 922.8093    &        $\rm ^{5}S\,^{0}_{2} - ^{5}P_{2}$      &  0.26     & 6.525      \\
1    & 923.7538    &        $\rm ^{5}S\,^{0}_{2} - ^{5}P_{1}$      &  0.04     & 6.525      \\ 
\hline
3    &1045.5449    &        $\rm ^{3}S\,^{0}_{1} - ^{3}P_{2}$      &  0.26     & 6.860      \\ 
3    &1045.6757    &        $\rm ^{3}S\,^{0}_{1} - ^{3}P_{0}$      & -0.43     & 6.860      \\
3    &1045.9406    &        $\rm ^{3}S\,^{0}_{1} - ^{3}P_{1}$      &  0.04     & 6.860      \\
\hline
6    & 869.3931    &4p-4d~~~$\rm ^{5}P_{3} - ^{5}D\,^{0}_{3}$      & -0.51     & 7.870      \\      
6    & 869.4626    &        $\rm ^{5}P_{3} - ^{5}D\,^{0}_{4}$      &  0.08     & 7.870      \\      
\hline
8    & 675.6851    &4p-5d~~~$\rm ^{5}P_{3} - ^{5}D\,^{0}_{2}$      & -1.76     & 7.870      \\
8    & 675.7007    &        $\rm ^{5}P_{3} - ^{5}D\,^{0}_{3}$      & -0.90     & 7.870      \\
8    & 675.7171    &        $\rm ^{5}P_{3} - ^{5}D\,^{0}_{4}$      & -0.31     & 7.870      \\
\hline
1F   &1082.1176    &        $\rm ^{3}P -^{1}D$                   & -8.62     & 0.000      \\ 
\hline
\end {tabular}  
\end {center}  
\end {table}

A rise in [S/Fe] with decreasing [Fe/H] is difficult to explain.  It
has been suggested that SNe with high explosion energies (hypernovae)
could have made a substantial contribution to the nucleosynthesis of
elements in the early Galaxy (Israelian \& Rebolo \cite{IR01},
Nakamura et al.  \cite{NUI01}), or even that they may be responsible for a
delayed deposition of the supernova-synthesized products into the
interstellar medium (Ramaty et al.  \cite{RSL00}).

On the other hand, Takeda et al.  (\cite{THT05}) noted that, even
taking into account the non-LTE effects on the formation of the S lines,
the resulting abundance of sulfur 
depended on the multiplet used for this determination.  
There was, in particular, an apparent discrepancy
between the abundance trends deduced from multiplets  6 and 1.

Taking advantage of a new model atom of sulfur (Korotin \cite{Kor08},
\cite{Kor09}), based on new radiative photoionization rates, we here
report S abundances for 33 very metal-poor stars, including 21
extremely metal-poor (EMP) stars with $\rm [Fe/H]<-2.9$.  Our
study is based on a precise high-resolution NLTE analysis of the lines
of the multiplet (hereafter Mult.)\,1 of \S1 \, at 
 $\lambda \lambda$ 921.29, 922.81, and 923.75~nm (see Table \ref{mults}).

\begin {figure}[hb]
\begin {center}
\resizebox{\hsize}{!}
{\includegraphics[clip=true]{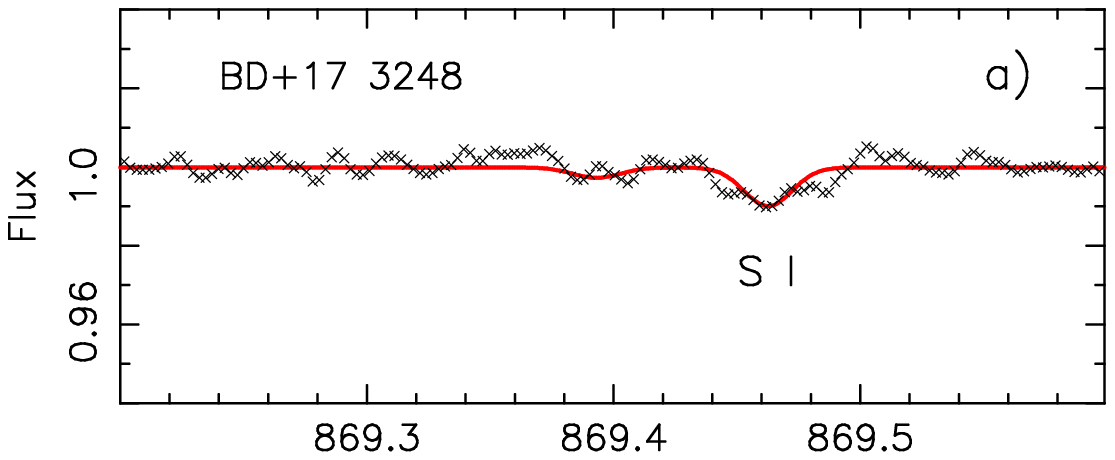}}
\resizebox{\hsize}{!}
{\includegraphics[clip=true]{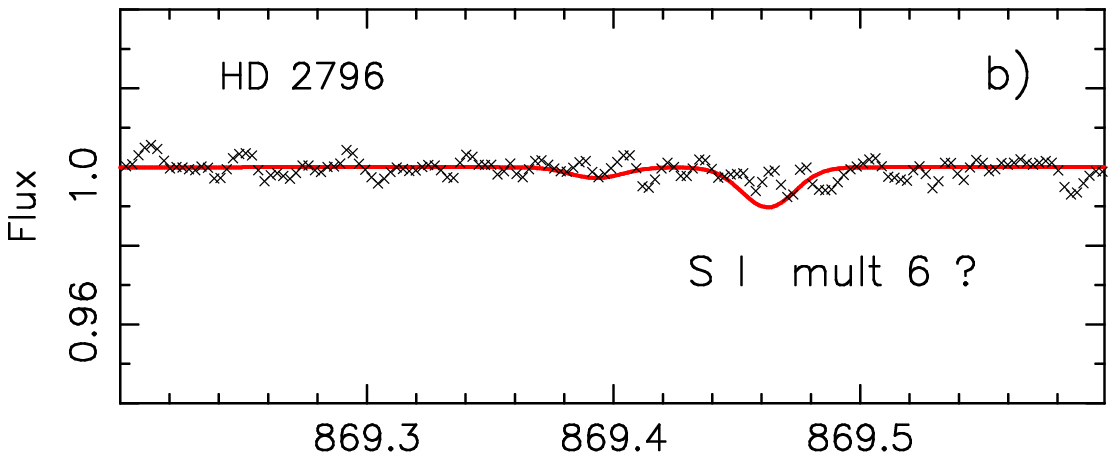}}
\caption{Spectra of BD+17\,3248 and HD\,2796, showing the line at
869.46~nm in Mult.\,6 of S\,{\small I}.  We detect the line in
BD+17\,3248 ([Fe/H]=--2.1), but not in HD\,2796 ([Fe/H]=--2.5).  A
synthetic spectrum (full line) corresponding to the equivalent width
of 0.26\,pm found by Israelian \& Rebolo for this line has been
superimposed on our spectrum of HD\,2796.}
\label{s8694}
\end {center}
\end {figure}

\section{Observations and reduction} 

All stars were observed as part of the ESO large programme ``First
Stars'', using the high-resolution spectrograph UVES (Dekker et al.
\cite{DDK00}) at the ESO-VLT. The spectra were discussed in
detail by Cayrel et al.  (\cite{CDS04}) and Bonifacio et al.
(\cite{BMS07}, \cite{BSC09}).  The resolving power of the spectra is
$R \approx 43\,000$ with 5 pixels per spectral resolution element.
The spectra were reduced using the UVES context within MIDAS
(Ballester et al.  \cite{BMB00}).  The signal-to-noise ratio (S/N) per pixel 
in the region of the S\,{\small I} lines is typically $\sim 180$ for the
giants and $\sim 90$ for the dwarfs (or 400 and 200 per resolution
element, see Cayrel et al., \cite{CDS04}).

Three Ba-poor giants (McWilliam et al.  \cite{MPS95}) were also
observed as part of the programme, under similar conditions as the
other metal-poor stars (see also Andrievsky et al.  \cite{ASK11}, to
be submitted  to A\&A).  The S abundance could be determined for one of
them (CS~22949-048), which has been added to the sample.

\section {Choice of sulfur lines and determination of fundamental
atmospheric parameters}
 In cool stars, five multiplets of \S1
can be observed in the red and near-infrared spectral region (Table
\ref{mults}): Mult.\,1 at about 922~nm, Mult.\,3 at 1045~nm, Mult.\,6
at 869~nm, and Mult.\,8 at 675~nm (Caffau et al.  \cite{CBF05},
\cite{CSF10}), and Mult.\,1F, the forbidden line at 1082.1 nm (Ryde,
\cite{Ryd06}, Caffau \& Ludwig, \cite{CL07}).  However, in EMP stars,
the lines of multiplets 6 and 8 are too weak to permit a precise analysis (see
Fig.~\ref{s8694}).  The lines of Mult.\,3 and Mult.\,1F, are 
non-detectable in stars with $\rm[Fe/H]<-2.9$, and are in any case outside
the range of our spectra.  The S abundances reported in this paper are
therefore derived from the lines of Mult.\,1 at 921.286, 922.809, and
923.754~nm.

\begin {figure}[h] \begin {center} 
\resizebox{\hsize}{!}
{\includegraphics[clip=true]{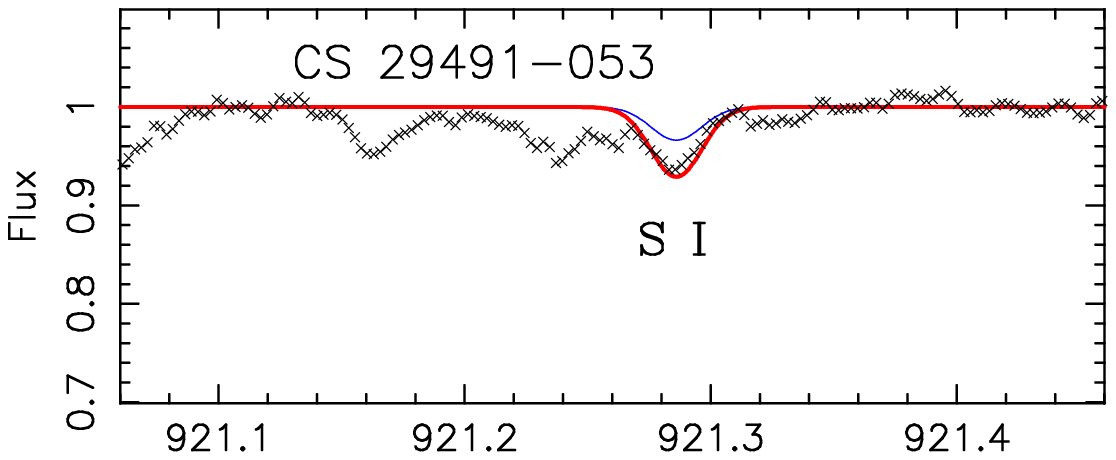}}
\resizebox{\hsize}{!}
{\includegraphics[clip=true]{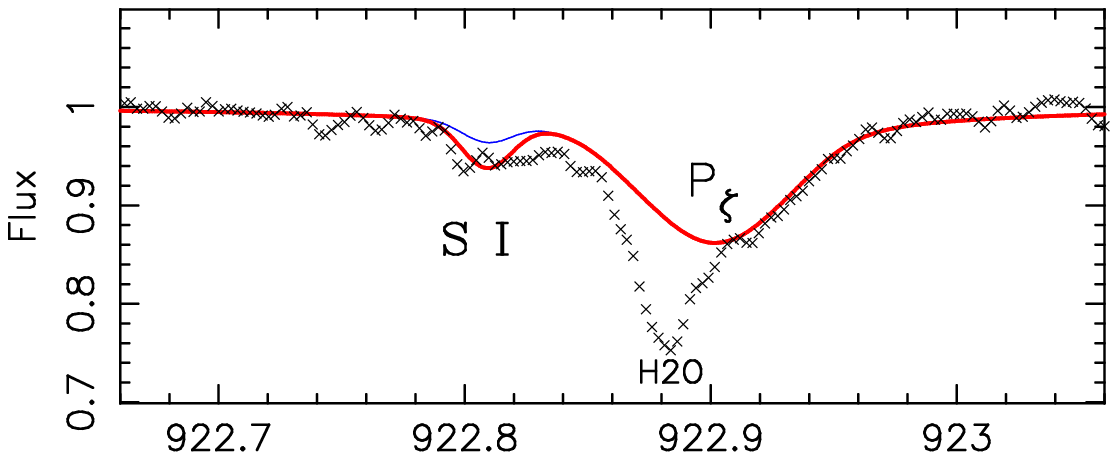}}
\resizebox{\hsize}{!}
{\includegraphics[clip=true]{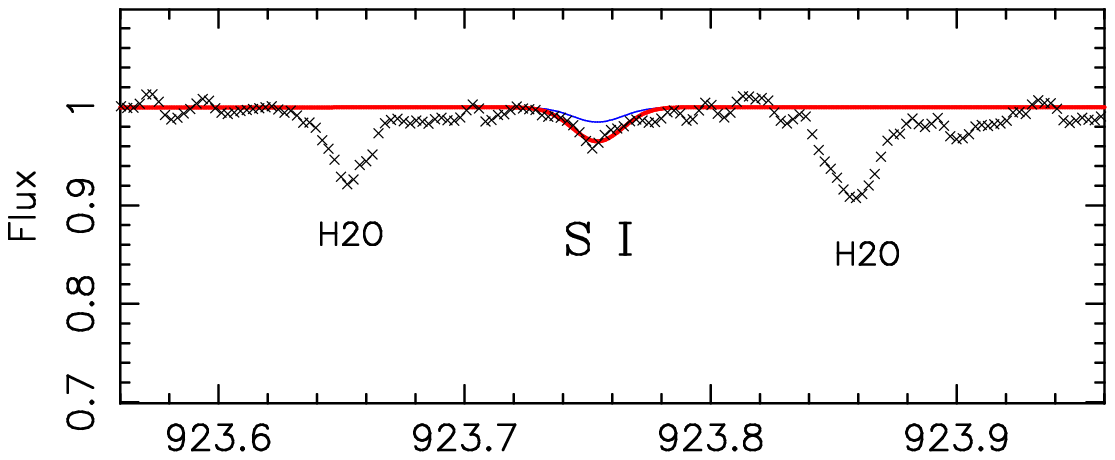}}
\caption{Spectrum of the giant star CS\,29491-053 ([Fe/H]=--3.0) in
the region of the \S1 lines of multiplet 1 (wavelength in nm).  The
synthetic spectrum (red line) has been computed for the adopted
abundance (Table \ref{tabstars}), and for an abundance 2.5 times
smaller (thin blue line).  The S line at 922.8 nm lies in the wing of
the hydrogen line $\rm P\zeta$; all other lines in this region are of
telluric origin.  In our analysis, all S lines blended with telluric
lines have been rejected.}
\label{spectra}
\end {center}
\end {figure}

\begin {table}[ht] 
\caption {Equivalent widths (pm) of the S\,{\small I} lines of multiplet 1.}  
\label {eqw} 
\begin{center}
\begin{tabular}{l r r r r }
Wavelength (nm)&  921.29 & 922.81 & 923.75   \\
$\rm \chi_{ex} (eV)$&  6.525  & 6.525  & 6.525   \\     
log $gf$ &    0.42   & 0.26   & 0.04   \\
&   EW (pm)&  EW (pm)&  EW (pm)   \\
\hline
{\bf Turnoff stars}\\ 
BS 16968-061&    1.68 &    -   &     -       \\
BS 17570-063&    1.30 &    -   &     -       \\
CS 22965-054&    1.30 &    -   &     -       \\
CS 29499-060&    1.80 &    -   &     -       \\
CS 29506-007&    1.98 &    -   &     -       \\
CS 29506-090&    1.63 &    -   &     -       \\
CS 30301-024&    1.95 &    -   &     -       \\
\hline   
{\bf Giants} \\
HD 2796     &    4.80 &   3.20 &   2.76      \\
HD 122563   &      -  &   1.92 &   1.29      \\
HD 186478   &      -  &   2.70 &   1.68      \\
BD+17\,3248 &      -  &   4.88 &             \\
BD-18~5550  &    1.65 &     -  &     -       \\
BS 17569-049&      -  &   1.20 &   1.20      \\
CS 22169-035&      -  &   1.00 &  (0.76)~\dag \\
CS 22186-025&   (2.09)&   1.52 &   0.79      \\
CS 22189-009&    0.72 &     -  &   0.32      \\
CS 22873-055&    2.37 &     -  &     -~      \\
CS 22873-166&    2.00 &   1.24 &   1.08      \\
CS 22878-101&      -  &     -  &   0.72      \\
CS 22891-209&    1.28 &  (1.00)&     -       \\
CS 22892-052&    1.18 &     -  &     -       \\
CS 22896-154&    1.98 &     -  &   0.79      \\
CS 22897-008&    0.90 &     -  &     -       \\
CS 22948-066&      -  &   1.20 &     -       \\
CS 22953-003&    1.43 &     -  &   0.74      \\
CS 22956-050&    1.08 &     -  &     -       \\
CS 22966-057&    2.49 &   1.04 &   1.15      \\
CS 29491-053&    1.53 &   1.02 &   0.88      \\
CS 29502-042&    0.83 &     -  &     -       \\
CS 29518-051&    2.30 &   1.34 &     -       \\
CS 30325-094&    1.33 &   0.98 &     -       \\
CS 31082-001&    2.19 &   1.90 &  0.97       \\
CS 22949-048&      -  &   0.55 &  0.45       \\
\hline
\multicolumn{5}{l}{\dag ~Equivalent widths in parenthesis are very uncertain}
\end {tabular}  
\end {center}  
\end {table}

The line at 922.809~nm is located in the wing of the Paschen $\zeta$
line, and the absorption from the hydrogen line has been taken into
account in the computations (see Fig.~\ref{spectra}).  This spectral
region is also crowded with telluric absorption lines of $\rm H_{2}O$.
Since no hot star was included among the programme stars, we
did not try to remove these telluric lines, but carefully
computed their position relative to the stellar lines using only 
S\,{\small I} lines free of contamination.  At least one or
two lines in each spectrum in general are free of blending.

Table \ref{eqw} provides equivalent widths of the \S1 ~lines, since
they were published neither in Cayrel et al.  (\cite{CDS04}) nor Bonifacio
et al.  (\cite{BSC09}).  In this spectral region, the observed spectra
are affected by fringing, which is not completely compensated when
dividing by a flat field (Nissen et al.  \cite{NAA07}).  The  continuum level
is therefore often not well defined, and we estimate
that the error in the equivalent width can reach 0.4\,pm for giants
and 0.5\,pm for dwarfs.  Since in all the cases, the sulfur lines in
Table \ref{eqw} are "weak'', the uncertainties, as a first
approximation, are  proportional to the EWs of the lines: $
\Delta(log~$\rm ab$) = \Delta(log~$\rm EW$)$.  However, these equivalent widths
were not used in the computations since we directly compared
synthetic and observed spectra; they are given here only for
comparison.

In total, we could thus determine S abundances in 26 giants and 7
turnoff stars from our original sample.

\begin {table*}[t] \caption {Adopted atmospheric parameters and NLTE S
abundances for our sample of EMP stars.  The solar abundance of
sulfur has been taken from Caffau et al.: $\rm log
\epsilon(S)_{\odot}=7.16$, the NLTE Mg abundance (column 6) is from
Andrievsky et al. (\cite{ASK10}).  The error (column 8) represents
the uncertainty in the spectral fitting.  The last column gives the
number of sulfur lines used in the determination } \label {tabstars}
\begin{center}
\begin{tabular}{l r r c r r r r r r r r r l}
~     &    ~        &     &           &      &      &                     &  ~  &               &     & 3D+  & 3D+   &3D+   \\
~     &    ~        &     &           &      & NLTE &        LTE          &err. &NLTE           & 3D  & NLTE & NLTE  &NLTE  \\
Star  name &$\rm T_{eff}$&log $g$&$\rm v_{t}$&[Fe/H]&[Mg/H]&$\rm log \epsilon(S)$&$\pm$&$\rm \log \epsilon(S)$&corr.&[S/H] & [S/Fe]&[S/Mg]&N\\
\hline
{\bf Turnoff stars}\\
\hline
BS~16968--061  & 6040&  3.8&  1.5&  -3.05& -2.46 & 4.90& 0.20 &4.33 &0.14 & -2.69& 0.36& -0.23&  1 \\
BS 17570--063  & 6240&  4.8&  0.5&  -2.92& -2.52 & 5.05& 0.20 &4.61 &0.12 & -2.43& 0.49&  0.09&  1 \\
CS~22965--054  & 6090&  3.8&  1.4&  -3.04& -2.42 & 4.74& 0.20 &4.21 &0.13 & -2.82& 0.22& -0.40&  1 \\
CS~29499--060  & 6320&  4.0&  1.5&  -2.70& -2.14 & 4.90& 0.20 &4.50 &0.12 & -2.54& 0.16& -0.40&  1 \\
CS 29506--007  & 6270&  4.0&  1.7&  -2.91& -2.25 & 5.03& 0.20 &4.62 &0.12 & -2.44& 0.47& -0.19&  1 \\
CS 29506--090  & 6300&  4.3&  1.4&  -2.83& -2.22 & 4.98& 0.20 &4.56 &0.12 & -2.48& 0.35& -0.26&  1 \\
CS 30301--024  & 6330&  4.0&  1.6&  -2.75& -2.25 & 4.98& 0.20 &4.60 &0.12 & -2.44& 0.31& -0.19&  1 \\
\hline
{\bf Giants}\\
\hline
HD 2796        & 4950&  1.5&  2.1&  -2.47& -1.84 & 5.25 &0.10 &4.97 &0.07 & -2.12& 0.35& -0.28& 3 \\
HD 122563      & 4600&  1.1&  2.0&  -2.82& -2.19 & 4.87 &0.10 &4.70 &0.06 & -2.40& 0.42& -0.21& 2 \\
HD 186478      & 4700&  1.3&  2.0&  -2.59& -1.86 & 5.07 &0.08 &4.84 &0.06 & -2.26& 0.33& -0.40& 2 \\
BD+17\,3248    & 5250&  1.4&  1.5&  -2.07& -1.39 & 5.57 &0.10 &5.11 &0.10 & -1.95& 0.12& -0.56& 2 \\
BD-18~5550     & 4750&  1.4&  1.8&  -3.06& -2.44 & 4.65 &0.12 &4.41 &0.07 & -2.68& 0.38& -0.24& 1 \\
BS 17569--049  & 4700&  1.2&  1.9&  -2.88& -2.11 & 4.70 &0.20 &4.40 &0.06 & -2.70& 0.18& -0.59& 2 \\
CS 22169--035  & 4700&  1.2&  2.2&  -3.04& -2.66 & 4.50 &0.11 &4.20 &0.05 & -2.91& 0.13& -0.25& 1 \\
CS 22186--025  & 4900&  1.5&  2.0&  -3.00& -2.39 & 4.70 &0.10 &4.43 &0.06 & -2.67& 0.33& -0.28& 2 \\
CS 22189--009  & 4900&  1.7&  1.9&  -3.49& -3.11 & 4.25 &0.20 &3.97 &0.07 & -3.12& 0.37& -0.01& 2 \\
CS 22873--055  & 4550&  0.7&  2.2&  -2.99& -2.34 & 4.72 &0.10 &4.52 &0.10 & -2.54& 0.45& -0.20& 1 \\
CS 22873--166  & 4550&  0.9&  2.1&  -2.97& -2.14 & 4.70 &0.10 &4.55 &0.07 & -2.54& 0.43& -0.40& 3 \\
CS 22878--101  & 4800&  1.3&  2.0&  -3.25& -2.56 & 4.55 &0.16 &4.36 &0.05 & -2.75& 0.50& -0.19& 1 \\
CS 22891--209  & 4700&  1.0&  2.1&  -3.29& -2.69 & 4.40 &0.15 &4.21 &0.06 & -2.89& 0.40& -0.20& 1 \\
CS 22892--052* & 4850&  1.6&  1.9&  -3.03& -2.52 & 4.48 &0.10 &4.29 &0.05 & -2.82& 0.21& -0.30& 1 \\
CS 22896--154  & 5250&  2.7&  1.2&  -2.69& -2.06 & 5.00 &0.08 &4.56 &0.07 & -2.53& 0.16& -0.47& 2 \\
CS 22897--008  & 4900&  1.7&  2.0&  -3.41& -2.83 & 4.35 &0.18 &4.10 &0.06 & -3.00& 0.41& -0.17& 1 \\
CS 22948--066  & 5100&  1.8&  2.0&  -3.14& -2.59 & 4.65 &0.14 &4.26 &0.06 & -2.84& 0.30& -0.25& 1 \\
CS 22949--037* & 4900&  1.5&  1.8&  -3.97& -2.42 &$<3.9$&  -  &$<3.6$&0.07&$<3.7$&$<0.5$&$<-1.1$& 2 \\
CS 22953--003  & 5100&  2.3&  1.7&  -2.84& -2.34 & 4.72 &0.10 &4.41 &0.06 & -2.69& 0.15& -0.35& 2 \\
CS 22956--050  & 4900&  1.7&  1.8&  -3.33& -2.59 & 4.45 &0.20 &4.18 &0.06 & -2.92& 0.41& -0.33& 1 \\
CS 22966--057  & 5300&  2.2&  1.4&  -2.62& -1.89 & 5.12 &0.10 &4.62 &0.08 & -2.46& 0.16& -0.57& 3 \\
CS 29491--053  & 4700&  1.3&  2.0&  -3.04& -2.34 & 4.60 &0.10 &4.45 &0.05 & -2.66& 0.38& -0.32& 3 \\
CS 29502--042  & 5100&  2.5&  1.5&  -3.19& -2.52 & 4.50 &0.14 &4.13 &0.06 & -2.97& 0.22& -0.45& 1 \\
CS 29518--051  & 5200&  2.6&  1.4&  -2.69& -2.02 & 4.95 &0.10 &4.50 &0.07 & -2.59& 0.10& -0.57& 2 \\
CS 30325--094  & 4950&  2.0&  1.5&  -3.30& -2.54 & 4.64 &0.18 &4.19 &0.06 & -2.91& 0.39& -0.37& 2 \\
CS 31082--001  & 4825&  1.5&  1.8&  -2.91& -2.09 & 4.84 &0.10 &4.54 &0.07 & -2.55& 0.36& -0.46& 3 \\
\\
CS 22949--048  & 4800&  1.5&  2.0&  -3.25& -2.68 & 4.42 &0.15 &4.07 &0.05 & -3.04& 0.21& -0.39& 2 \\
\hline   
\end {tabular}  
\end {center}  
* An asterisk after the name of the star means that the star is carbon-rich.   
\end {table*}

\section{NLTE calculations and 3D correction}

The atmospheric parameters of the stars (\Teff, \logg, [Fe/H]) were adopted
from Cayrel et al.  (\cite{CDS04}), Hill et al.  (\cite{HPC02}), and
Bonifacio et al.  (\cite{BMS07}) and for the Ba-poor star  from
Andrievsky et al.  (\cite{ASK11}).  To summarize, the temperature of the
giants were inferred from their colours by adopting the calibration
of Alonso et al.  (\cite {AAM96}), and the temperature of the turnoff
stars from the wings of the $\rm H_{\alpha}$ line.  The random error in the
temperature is about 80\,K ($1 \sigma$).  For a given stellar
temperature, the ionization equilibrium provides an estimate of the
stellar gravity with an internal accuracy of about 0.1~dex in \logg,
and the microturbulence velocity \vt~can be constrained to within $\rm
0.2 ~km s^{-1}$.  The adopted parameters are given in Table
\ref{tabstars}.  Table \ref{errTgv} lists the uncertainty in the
sulfur abundance originating in the errors in \Teff, \logg~, and \vt.
Because gravity is determined from the ionization equilibrium, a
variation in \Teff ~will change \logg~ and also sometimes slightly
influences \vt.  Hence, the total error budget is not the quadratic
sum of the various sources of uncertainties, but contains significant
covariance terms (see Cayrel et al.  \cite{CDS04} for more details).

Our \logg~value may be affected by NLTE effects (overionization) and
the uncertainties in the oscillator strengths of the Fe and Ti lines.
This was discussed in Cayrel et al.  (\cite{CDS04}), their section 3.1.
 : even for giants the spectroscopic \logg~ and
the \logg~ value deduced from isochrones should not differ by more
than about 0.3~dex.

\begin {table}[t]
\caption {Abundance uncertainties linked to stellar parameters.
Case of a typical dwarf CS~29506-090, and a typical giant HD~122563.}
\label {errTgv}
\begin {center}
\begin {tabular}{lrrrr}
\hline
\multicolumn {2}{l}{\bf CS~29506-090}\\
\multicolumn {5}{c}{A: \Teff=6300K, \logg=4.3 dex, vt=1.4 
km s$^{-1}$}\\
\multicolumn {5}{c}{B: \Teff=6300K, \logg=4.2 dex, vt=1.4 
km s$^{-1}$}\\
\multicolumn {5}{c}{C: \Teff=6300K, \logg=4.3 dex, vt=1.2 
km s$^{-1}$}\\
\multicolumn {5}{c}{D: \Teff=6200K, \logg=4.3 dex, vt=1.4 
km s$^{-1}$}\\
\multicolumn {5}{c}{E: \Teff=6200K, \logg=4.1 dex, vt=1.3 
km s$^{-1}$}\\
\hline
El.   & $\Delta_{B-A} $ & $\Delta_{C-A} $& $\Delta_{D-A} $& 
$\Delta_{E-A} $\\
\hline
[Fe/H]      &-0.03 & 0.03 &-0.06 &-0.07\\    
$[$S I/Fe]  & 0.00 &-0.03 & 0.09 & 0.04\\
$[$Fe I/Fe] & 0.02 & 0.01 &-0.04 & 0.01\\
$[$Fe II/Fe]&-0.01 &-0.01 & 0.04 &-0.01\\
\hline
\hline
\multicolumn {2}{l}{\bf HD~122563}\\
\multicolumn {5}{c}{A: \Teff=4600K, \logg=1.0 dex, vt=2.0 
km s$^{-1}$}\\
\multicolumn {5}{c}{B: \Teff=4600K, \logg=0.9 dex, vt=2.0 
km s$^{-1}$}\\
\multicolumn {5}{c}{C: \Teff=4600K, \logg=1.0 dex, vt=1.8 
km s$^{-1}$}\\
\multicolumn {5}{c}{D: \Teff=4500K, \logg=1.0 dex, vt=2.0 
km s$^{-1}$}\\
\multicolumn {5}{c}{E: \Teff=4500K, \logg=0.6 dex, vt=1.8 
km s$^{-1}$}\\
\hline
El.   & $\Delta_{B-A} $ & $\Delta_{C-A} $& $\Delta_{D-A} $& 
$\Delta_{E-A} $\\
\hline
[Fe/H]      &-0.00 & 0.06 &-0.09 &-0.06\\    
$[$S I/Fe]  &-0.02 &-0.02 & 0.12 & 0.04\\
$[$Fe I/Fe] & 0.03 & 0.03 &-0.11 & 0.03\\
$[$Fe II/Fe]&-0.03 &-0.04 & 0.11 &-0.03\\
\hline
\end {tabular}
\end {center}
\end {table}

For the non-LTE computations of the sulfur abundance, Kurucz models
without overshooting were used (Castelli et al., \cite{CGK97}).  These
models have been shown to provide LTE abundances very similar (within
0.05~dex) to those of the MARCS models used by Cayrel et al.
(\cite{CDS04}) and Bonifacio et al.  (\cite{BSC09}).

\begin {figure}[h]
\begin {center}
\resizebox{4.3cm}{4.3cm}
{\includegraphics[clip=true]{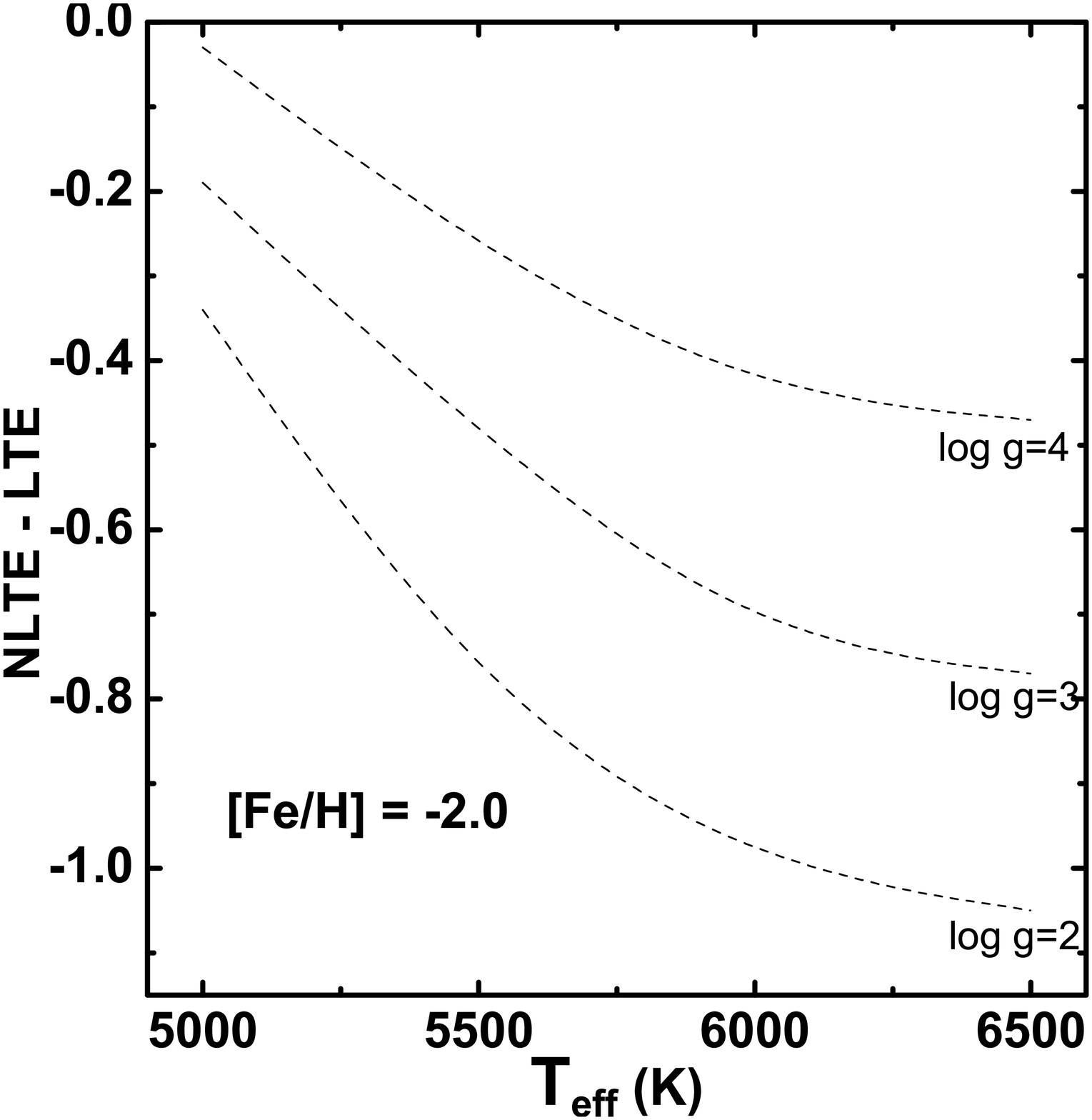}}
\resizebox{4.3cm}{4.3cm} 
{\includegraphics[clip=true]{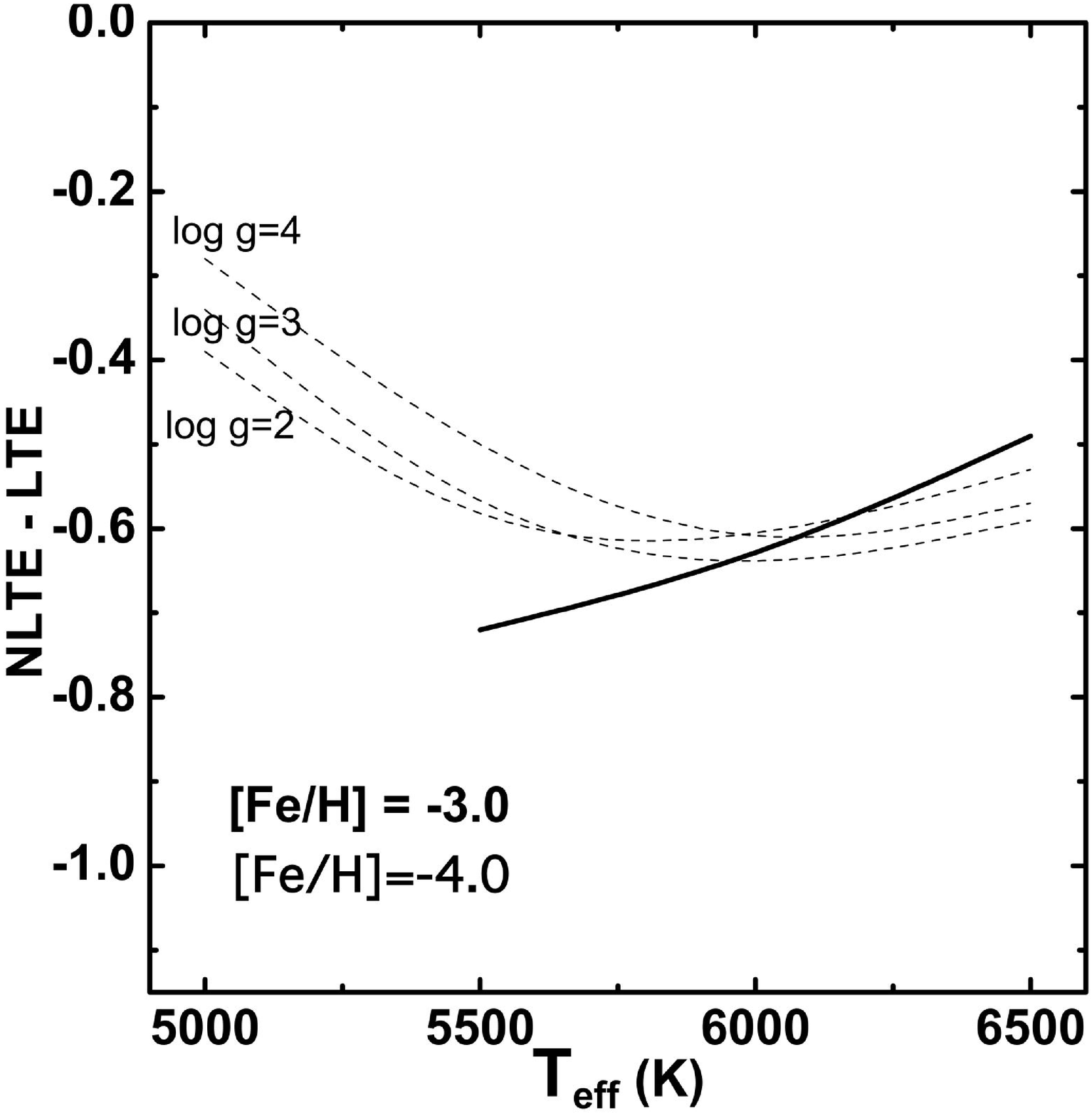}}
\caption{NLTE correction for the 921.2nm sulfur line in metal-poor
stars as a function of the temperature, the gravity, and the
metallicity for [Fe/H]=--2, [Fe/H]=--3 (dashed lines) and [Fe/H]=--4
(solid line).  This correction has been computed for [S/H]=[Fe/H].  }
\label{cornlte}
\end {center}
\end {figure}

It has been shown (see Takeda et al.  \cite{THT05}, Korotin
\cite{Kor08}) that Mult.\,1 of S\,{\small I} is affected by a
significant negative non-LTE correction.  To compute NLTE profiles of
the S\,{\small I} lines, we used the model atom of Korotin
(\cite{Kor08}), which contains 64 levels of S\,{\small I} and the
ground level of S\,{\small II}.  The radiative photoionization rates
of all the levels taken into account are based on the new detailed
ionization cross-sections listed in the Opacity Project TopBase
(hereafter TopBase)\footnote{http://cdsweb.u-strasbg.fr/topbase/topbase.html}.  
The oscillator
strengths of the lines are also taken from TopBase.  The line profiles
were computed from a modified version of the program MULTI
(Carlsson \cite{Car86}, Korotin et al.  \cite{KAL99}).  This version
includes, in particular, opacity sources from ATLAS9 (Kurucz
\cite{Kur92}), which modify the continuum opacity and the intensity
distribution in the UV region.  These sources are very important for
determining of the radiative rates of the bound-bound transitions in
the sulfur atom.  In Fig.  \ref{cornlte}, we present the new non-LTE
correction computed for metal-poor stars for the 921.2~nm sulfur line
of the Mult.\,1.  This correction is sometimes rather different from
the correction computed by Takeda et al.  (\cite {THT05}).  For
example, for turnoff stars with a metallicity of about --3.0 and a
temperature of 6000\,K, the correction is found to be close to
--0.6~dex, whereas it was only about --0.2~dex following Takeda et al.
(\cite {THT05}).

We also computed the 3D correction ({as explained in Caffau et
al.  \cite{CLS10} or Caffau \& Ludwig \cite{CL07}}).  This correction
is small, but positive (Table \ref{tabstars}), and has been computed
and applied separately.

\subsection{Test of coherence}
With the model atom adopted here, there is good agreement (to within 0.1
dex) in the Sun and in Procyon between the S abundances derived from
the different multiplets when 3D models are used.
For Procyon, we found $\rm \epsilon (S)=7.23\pm 0.03$ from the 3D,
LTE computations of Caffau et al.  (\cite{CFB07}) with a Korotin's
NLTE correction (private communication).
For the Sun, the mean sulfur abundance derived from these
computations is log $\rm \epsilon (S)=7.16\pm 0.05$, for log $\rm
\epsilon (H)=12.0$ (Caffau et al.  (\cite{CLS10}).  This value is very
close to the meteoritic value ($7.17\pm 0.02$, following Lodders et
al., \cite{LPG09}).  The solar abundance $\rm log \epsilon (S)=7.16$
has been adopted as a reference in our computations.  \\
Moreover, for the most metal-rich star of our sample, BD+17\,3248, the
S abundance could be determined from the lines of the Mult.\,1 and the
main line of Mult.\,6 (Fig.~\ref{s8694}a).  The agreement between the
NLTE S abundances derived from multiplets 6 and 1 is very good when
the 3D correction is also included (-0.02 for Mult.\,6, +0.10 for
Mult.\,1): $\rm \epsilon(S)=5.28\pm 0.2$ from Mult.\,6 and $\rm
\epsilon(S)=5.22\pm 0.1$ from Mult.\,1.

\section{Results}

In Table \ref{tabstars}, we present the result of the computations for
our sample of stars.  The S abundances were derived from fits of
synthetic spectra (Fig.~\ref{spectra}).  The sulfur abundances
could not be determined for the most metal-poor stars ($\rm [Fe/H] <
-3.6$ for giants, $\rm[Fe/H] < -3.2$ for dwarfs), because the lines
were too weak to be measured reliably.  However, an upper limit to the
sulfur abundance was given for CS~22949-037 because this upper
limit can help us to explain the chemical anomalies of this star (see
section \ref{star22949-37}).  The error in the sulfur abundance,
given in Table \ref{tabstars}, represents the uncertainty in the
spectral fitting.  This error strongly dominates the total error in both
[S/Fe] and [S/Mg].

\subsection{[S/Fe] vs. [Fe/H] and comparison to previous studies}
\begin {figure}[t]
\begin {center}
\resizebox{\hsize}{!}
{\includegraphics[clip=true]{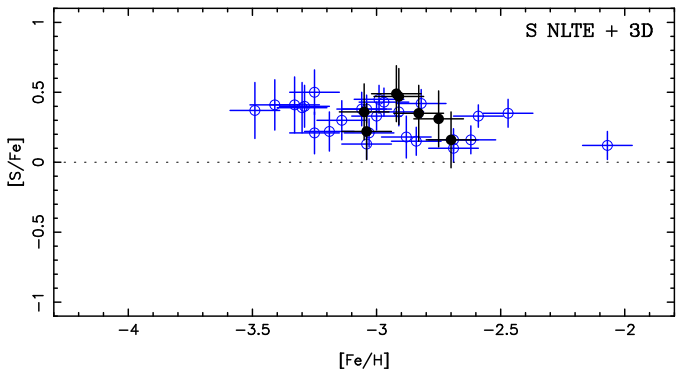}}
\caption{Final NLTE [S/Fe] ratios for our EMP stars.  Solid symbols
show the turnoff stars, open symbols the giants.}
\label {S-Fe}
\end {center}
\end {figure}

Fig.~\ref{S-Fe} shows [S/Fe] as a function of [Fe/H] for our sample of
EMP stars.  Below [Fe/H]=--2.9, the slope of the relation is very
small :  $\rm a=-0.012 \pm 0.006$,  thus [S/Fe] is practically constant
with a mean of $\rm \overline{[S/Fe]}=+0.35 \pm 0.10$.  The level of
this plateau is in good agreement with the value found by Nissen et
al.  (\cite{NCA04}, \cite{NAA07}) and Caffau et al.  (\cite{CBF05}),
$\rm \overline{[S/Fe]} \approx +0.35 $ (found for the most metal-poor stars
of their sample).  \\


\begin {figure}[hb] 
\begin {center} 
\resizebox{\hsize}{!}
{\includegraphics[clip=true]{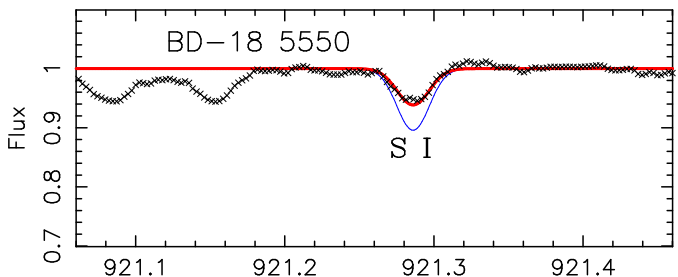}}
\caption{ Spectrum of
BD--18~5550 in the region of Mult.\,1 of S\,{\small I}.  Only the main
line can be measured in our spectrum, as the others are severely
blended with telluric lines.  The thick line shows the theoretical
profile computed with our value of the S abundance (Table
\ref{tabstars}).  The profile computed with the abundance adopted by
Takeda \& Takada-Hidai (2010) (thin line), is not compatible with the
observed spectrum.} 
\label {sp5550} 
\end {center} 
\end {figure}

\noindent $\bullet$ {\bf Behavior of stars with $\rm -2.5 <[Fe/H]$}\\
It is interesting that none of the stars in Fig.~\ref{S-Fe} is found
to be "S-rich'' ($\rm [S/Fe]>+0.7$), in contrast to several stars in
the interval $\rm -2<[Fe/H]<-1 $ (Israelian \& Rebolo \cite{IR01},
Takada-Hidai et al.  \cite {TTS02}, Takeda et al.  \cite {THT05}, and
Caffau et al.  \cite{CBF05}).

The most S-rich star in Israelian \& Rebolo (\cite{IR01}) is HD\,2796,
for which [S/Fe]=+0.8. It is included in our sample, but its very
high S abundance is not confirmed by our measurement of the lines of
Mult.\,1 from which we find [S/Fe]=+0.3.  Israelian \& Rebolo found 
that EW=0.26\,pm for the main line of Mult.\,6 in this star, 
but although our spectrum
has a very high S/N in this region (over 500/pix at 869 nm), we do not
detect the line (Fig.~\ref{s8694}b).  Its equivalent width is clearly
smaller than 0.26\,pm,  the error in the equivalent width 
was  probably underestimated by Israelian \& Rebolo.  Ryde
\& Lambert (\cite{RL04}), reanalysed three other S-rich stars from
the list of Israelian \& Rebolo with $\rm 0.55 <[S/Fe]< 0.7$ and for
all of them they found also found a normal sulfur abundance.

In Caffau et al.  (\cite{CBF05}), three other stars were found to
be S-rich (with $\rm [S/Fe]>+0.7$), BD+02\,263, BD+05\,3640 and
CD-45\,13178 (HD\,181743), their metallicities being in the
interval $\rm -2.2<[Fe/H]<-1.1 $, thus outside the range of
metallicity explored in our paper.  Moreover, in HD 181743, Nissen et
al.  (\cite{NAA07}) found from the Mult.  \,1 a normal abundance of
sulfur ([S/Fe]=+0.34), but on the other hand, in BD+05\,3640 the high
sulfur abundance has been confirmed (but at a lower level) by Caffau
et al.  (\cite {CSF10}) by the analysis of the Mult.\,3.

It remains unclear wether S-rich stars exist with metallicities
in the interval $\rm-2.5 <[Fe/H] < -1.1$.\\

\noindent $\bullet$ {\bf Behavior of stars with $\rm [Fe/H]<-2.5$}\\
Recently, Takeda \& Takada-Hidai (\cite{TT10})  measured the Mult.\,3
of S\,{\small I} in metal-poor stars and found [S/Fe] to be constant in
the interval $\rm -2.5 <[Fe/H] < -1.1$, but to increase suddenly for
$\rm [Fe/H] < -2.6$.  They have only five stars with $\rm [Fe/H] <
-2.6$, and in one of them the S line cannot be detected.  For G64-37,
the uncertainty is so high that the detection is in doubt (see their
Fig.  13).  Of the three remaining stars, two of them are in common
with our sample: HD~122563 and BD--18~5550.  For these stars, we find
"normal'' S abundances of [S/Fe]=+0.42 and +0.38 from Mult.\,1 (Table
\ref{tabstars}).

To understand the cause of this discrepancy, we remeasured the
equivalent widths of the 1045.5449~nm line in the spectra of Takeda \&
Takada-Hidai (\cite{TT10}), neglecting the other two lines, which are
even fainter.  The lines of Mult.\,3 used by Takeda \& Takada-Hidai
are weaker than those of Mult.\,1 and their spectra have only half the
resolution of ours. For HD\,122563, we find that EW=1.5~pm, 
corresponding to [S/Fe]= +0.45,
in excellent agreement with our determination. Since the sulfur
lines are weak, the value of [S/Fe] found by Takeda \& Takada-Hidai was
simply corrected by $\rm \Delta [S/Fe]= log(EW_{2}/EW_{1}$),
 where EW1 and EW2 represent the old and new equivalent widths.
BD--18~5550 is the most S-rich star in the sample of Takeda \&
Takada-Hidai with [S/Fe]=+0.76, against our value of +0.38 from
Mult.\,1 (see Fig.  \ref{sp5550}).  For this star, the S/N of the
spectrum of Takeda \& Takada-Hidai seems to be rather high, but we
note that the full width at half maximum (FWHM) of the lines 
is much larger in this spectrum than in 
the spectrum of HD\,122563, while in our UVES spectra
the FWHM of the lines in both stars 
is about the same. If we fix the FWHM of
the lines in BD--18~5550 to the same value as in HD~122563, we find
the equivalent width of the 1045.5449~nm line to be about 1.2~pm and
[S/Fe]=+0.62.  Adopting the Cayrel formula (Cayrel \cite{Cay88}) to
estimate the precision of this measurement (with S/N =150), we find
the error of this equivalent width to be close to 0.5~pm.  Hence, the
error in the abundance ratio should be about 0.2~dex, which makes
their result compatible with ours.


\begin {figure}[ht]
\begin {center}
\resizebox{\hsize}{!}
{\includegraphics[clip=true]{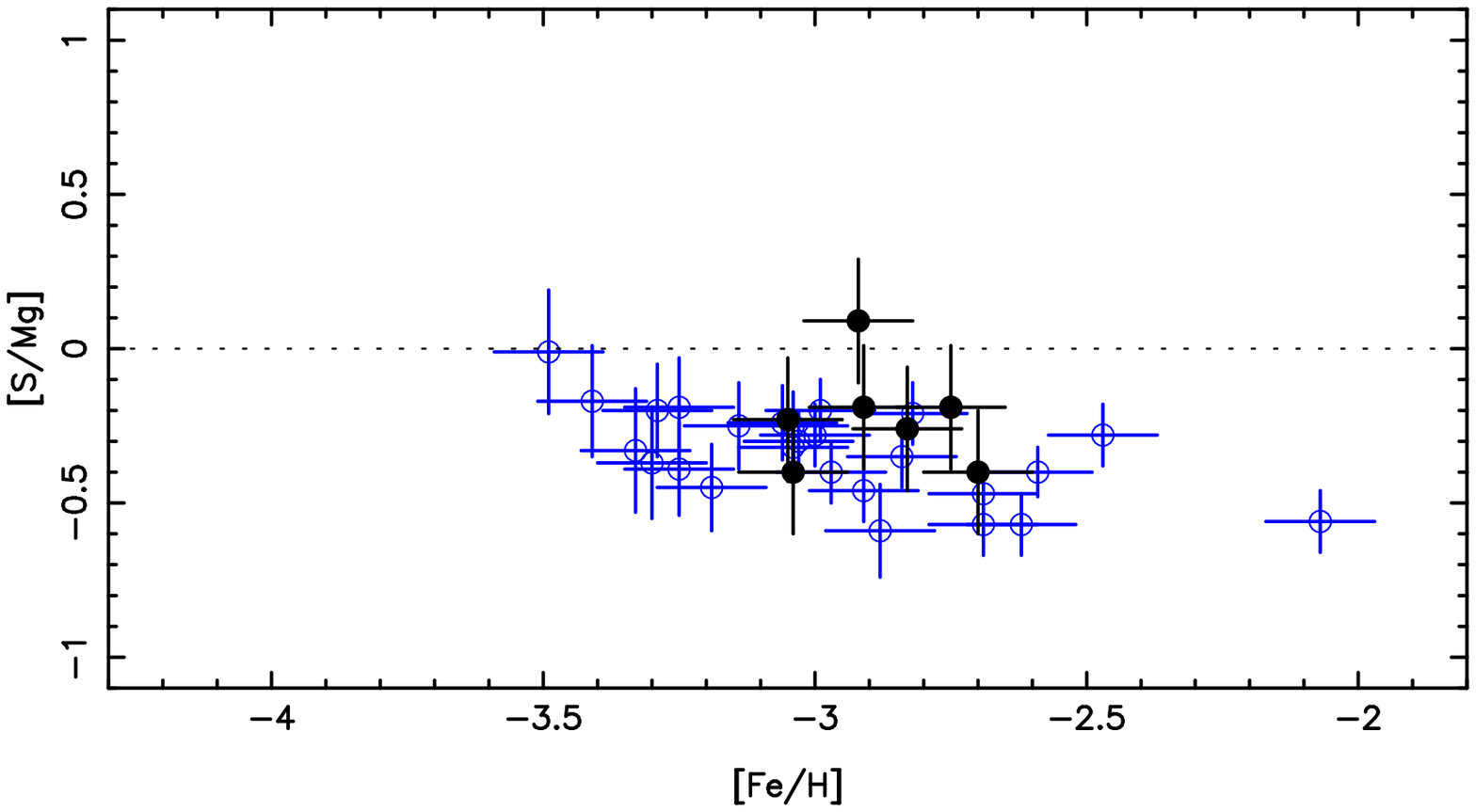}}
\resizebox{\hsize}{!}
{\includegraphics[clip=true]{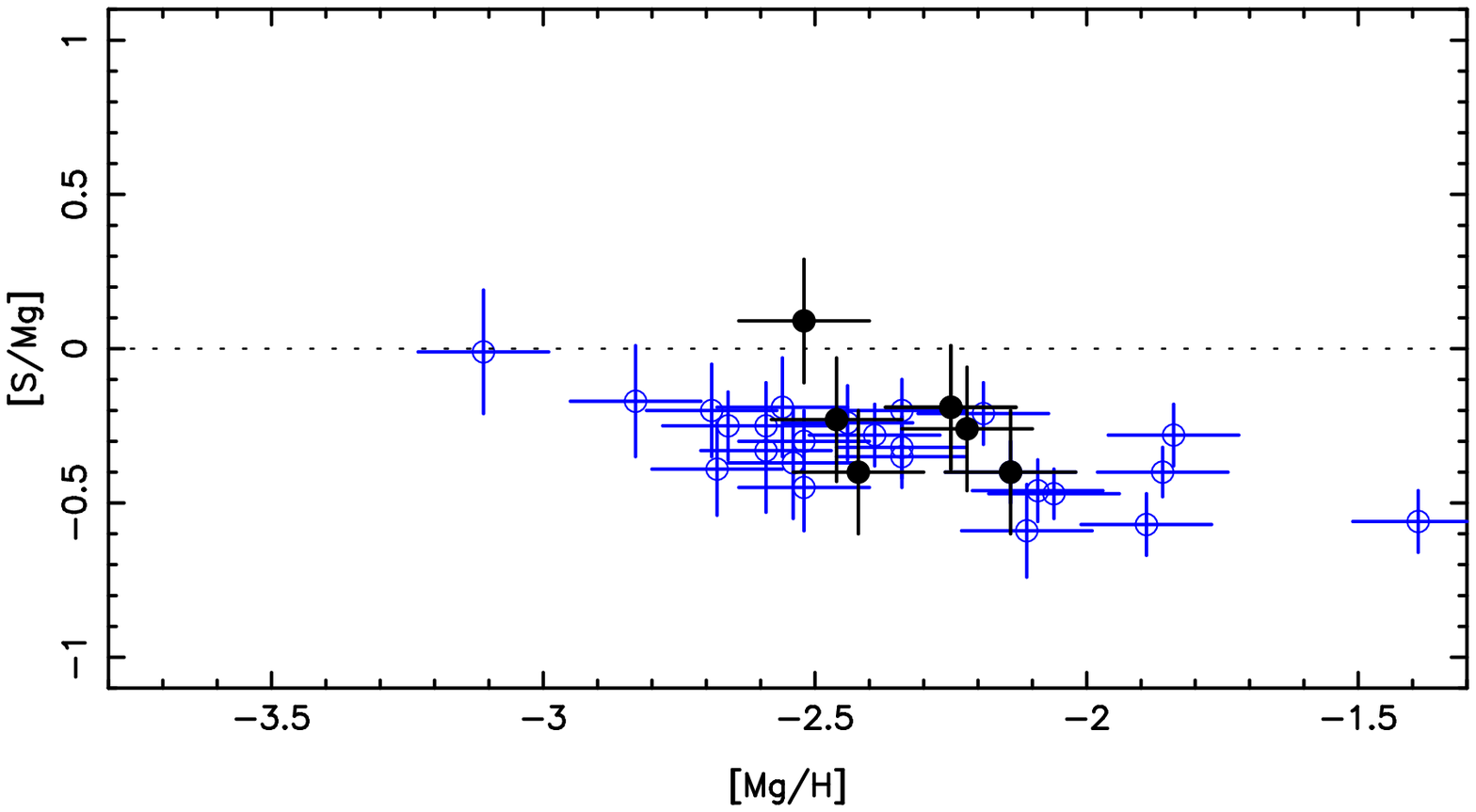}}
\caption{
[S/Mg] vs. [Fe/H] (top) and [Mg/H] (bottom). Symbols as in Fig. \ref{S-Fe}.}
\label {smg}
\end {center}
\end {figure}

\subsection{[S/Mg] vs. predicted supernova yields}

It is often noted that magnesium should be a more reliable reference element
than iron, because Mg is only formed in massive SN II and its
production is less affected by explosive nucleosynthesis, mixing, and
"fallback''.  As a consequence, the predictions for [S/Mg] should be
more robust than for [S/Fe] (Woosley \& Weaver, \cite {WW95},
Shigeyama \& Tsujimoto \cite{ST98}).

The NLTE abundance of Mg for our sample of stars was computed by
Andrievsky et al.  (\cite{ASK10}).  Fig.~\ref{smg} shows the observed
value of [S/Mg] as a function of [Fe/H] and [Mg/H].  The scatter is
small, but two stars seem to have a high [S/Mg] ratio: BS~17570--063
([Fe/H]=--2.92) and CS~22189--009 ([Fe/H]=--3.49).  In both cases, the
high [S/Mg] ratio seems to be real, and not the consequence of an
uncertainty in the magnesium abundance (see Fig.  \ref{smg}). 
The full spread of NLTE [Mg/Fe] for the metal-poor stars below $\rm
[Fe/H]<-1$ is definitely larger than expected from the error estimate
of the analysis itself (see Gehren et al., {\cite{GSZ06}}, Andrievsky
et al., \cite{ASK10}, and also Cayrel et al., \cite{CDS04}) and the
anomalies of the magnesium abundances are not clearly correlated with
the other elements.

In the interval $\rm -3.4 < [Fe/H] <-2.5$, we can define a mean value
of [S/Mg] as done for O, Na, Al and K by Andrievsky et al.
(\cite{ASK10}): $\rm \overline{[S/Mg]}=-0.32 \pm 0.14$.  In the same
metallicity interval, the scatter in [S/Fe] around the mean is only
0.12~dex (see Fig \ref{S-Fe}). It appears that, as already observed
in LTE (Cayrel et al.  \cite{CDS04}), the scatter of the abundance
ratios is slightly larger when Mg rather than Fe is used as the
reference element.  This is surprising because Fe forms in
processes quite different from those forming Mg and S (explosive
burning for Fe, and mainly hydrostatic C, Ne, and O burning for Mg and
S).

For $\rm [Mg/H]<-2.7$, a slight increase in [Na/Mg], [Al/Mg], and
[K/Mg] as [Mg/H] decreases was seen by Andrievsky et al.
(\cite{ASK10}).  A slight increase in [S/Mg] is also possible (Fig.
\ref{smg}), but there are very few stars with $\rm [Mg/H] < -2.7$ in
our sample.  We plan to discuss the evolution of [S/Mg] in the Galaxy
in more detail when we have completed a new precise analysis of disk
stars, including NLTE and 3D corrections.

\begin {figure}[htb]
\begin {center}
\resizebox{7.0cm}{6.0cm}
{\includegraphics[clip=true]{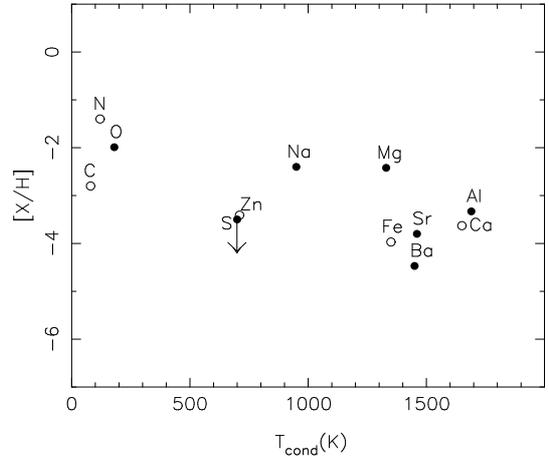}}
\caption{ [X/H]
abundances in CS~22949-037 against the solar condensation temperatures
(Lodders, \cite{Lod03}).  There is no correlation between $\rm
T_{cond}$ and the abundance of the elements, in particular S and Zn
have a condensation temperature lower than Na and they are more
deficient, at variance with the abundances in $\rm\lambda ~Boo$ stars.
(The filled circles represent the abundances corrected or not
sensitive to NLTE effects, the open symbols are abundances that could
not be corrected for NLTE effects) } 
\label {22949a} 
\end {center}
\end {figure}

\subsection{The case of the peculiar C-rich star CS~22949-037} \label{star22949-37}
CS~22949-037 is a very peculiar giant: it is extremely metal-poor
($\rm [Fe/H]\approx -4.0$), and its unusual chemical composition was
first pointed out by McWilliam et al.  (\cite{MPS95}) and later
studied in more detail by Norris et al.  (\cite{NRB01}) and Depagne et
al.  (\cite{DHS02}).  This star exhibits light element (C-N-O, Na, Mg)
enhancements that are quite similar to those associated with the hyper
iron-poor stars HE 0107-5240 and HE 1327-2326 with $\rm[Fe/H]<-5$
(see Frebel et al., \cite{FAC05}, \cite{FCE08}).  It is generally
admitted that these stars are the product of specific zero-metal
supernovae.  However, Venn \& Lambert (\cite{VL08}) argued that
this abundance pattern might be associated with the same peculiarity
that is used to account for $\rm \lambda Boo$ stars (presumably
related to dust depletion).  If this were so, the abundance of the
elements should be smoothly correlated with the condensation
temperature $\rm T_{cond}$.  The abundances of S and Zn could help us to
choose between the two hypotheses, because their temperatures are
intermediate between those of O and Na, both of which are strongly
enhanced in CS~22949-037.  If dust depletion is responsible for the
pattern of CS~22949-037 we would expect that sulfur and zinc also
be strongly enhanced in this star.  The abundance of zinc was 
measured in CS~22949-037 by Depagne et al.  (\cite{DHS02}).  The
sulfur lines are not visible in this star but 
(as can be seen in Table \ref{tabstars})
an upper limit can be estimated as $\rm [S/H]<-3.5$ or $\rm
[S/Fe]<+0.5$.  In Fig.  \ref{22949a}, we have plotted [X/H] vs.
 $\rm T_{cond}$.  The abundances [X/H] of C, N, O, Ca, Fe, and Zn are taken
from Depagne et al. (\cite{DHS02}); for the other elements Na, Mg,
Al, S, Sr, and Ba, the NLTE abundances
(Andrievsky et al., \cite{ASK07}, \cite{ASK08}, \cite{ASK09},
\cite{ASK10}, \cite{ASK11})  have been preferred.  The oxygen
abundance, determined from the forbidden line, is insensitive to the
NLTE effects.  It can be seen that CS~22949-037 does not show any
clear signature of dust depletion.

\subsection{Comparison with predicted SN yields}

\begin {figure}[ht]
\begin {center}
\resizebox {4.3cm}{2.0cm}
{\includegraphics {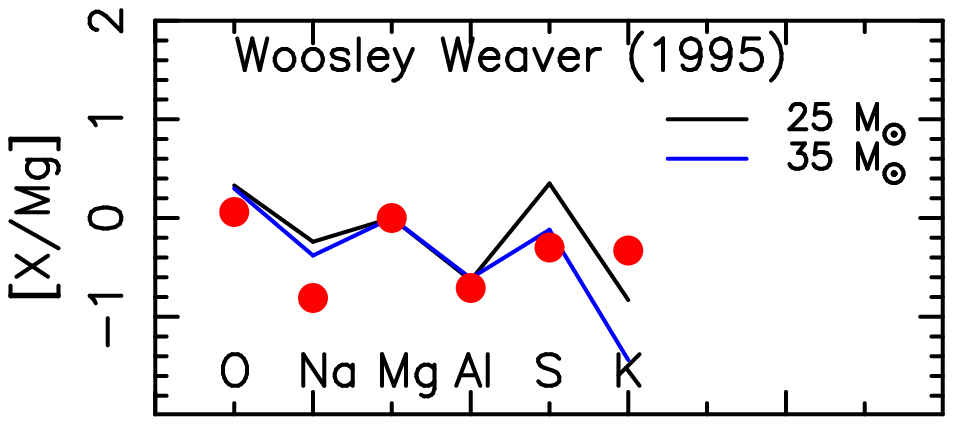} }
\resizebox  {4.3cm}{2.0cm}
{\includegraphics {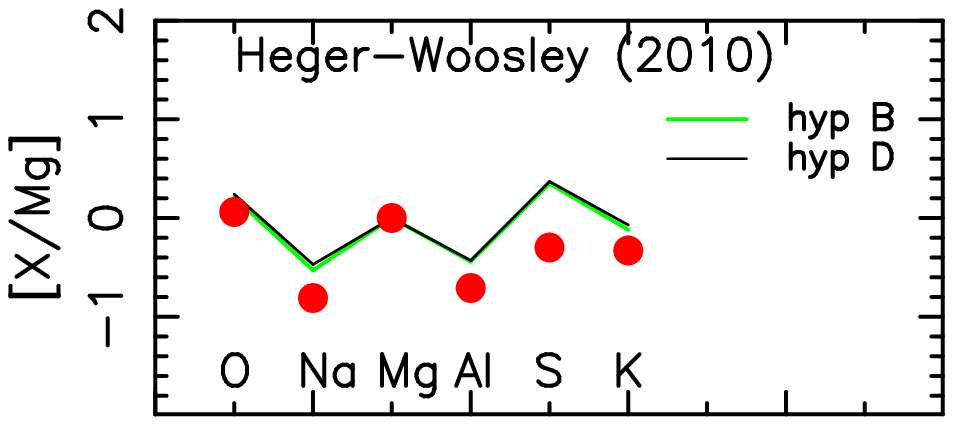} }
\resizebox  {4.3cm}{2.0cm}
{\includegraphics {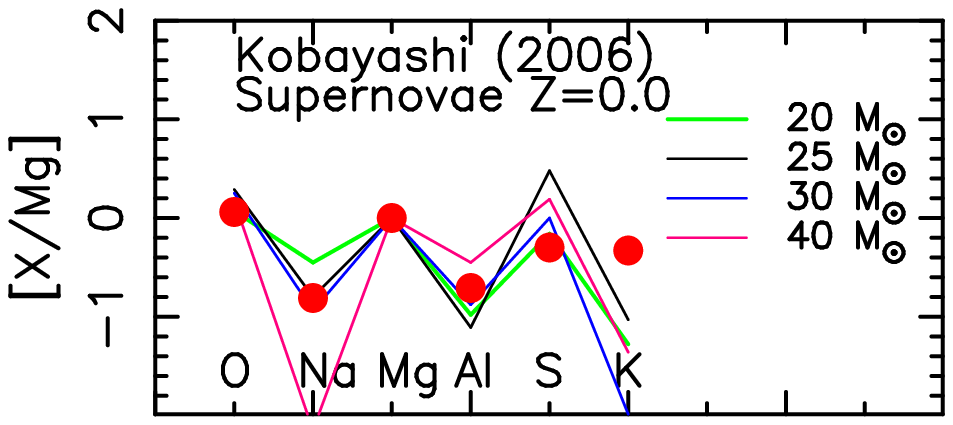} }
\resizebox  {4.3cm}{2.0cm}
{\includegraphics {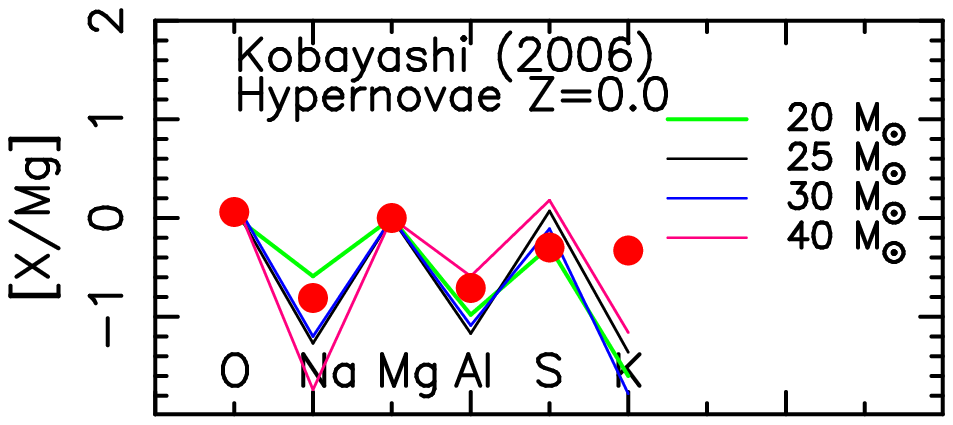} }
\caption{ Comparison of the most recent O, Na Mg, Al, S, and K
abundances in EMP stars to the predicted yields of metal-poor SNe and
hypernovae.  The solid lines represent the predictions of different
models, and the dots the mean values of the observed abundances in the
interval $\rm -3.0<[Mg/H]<-2.0$}
\label {SN}
\end {center}
\end {figure}

In Fig.  \ref{SN}, we compare the abundance ratios of the light metals
to some predicted abundances in the ejecta of metal-poor SNe or
hypernovae following Woosley \& Weaver (\cite{WW95}), Kobayashi et al.
(\cite{KUN06}), and Heger \& Woosley (\cite{HW10}). The mean values of
the observed abundances of the different elements are computed in the
interval $\rm -3.0<[Mg/H]<-2.0$, their uncertainty being of the
order of 0.1 dex.  The amount of ejected S seems to be generally
overestimated by the models.  If we consider all the elements from O
to K, the best agreement is obtained with the predictions by the
models of Heger \& Woosley \cite{HW10} (hypotheses B or D, with a low
mass cutoff at 10$M_{\odot}$), mainly because their predictions show
a better agreement with the observed K abundance.

\subsection{[S/Zn] in metal-poor stars and damped $\rm Ly\alpha$
systems}

\begin {figure}[ht]
\begin {center}
\resizebox{\hsize}{!}
{\includegraphics[clip=true]{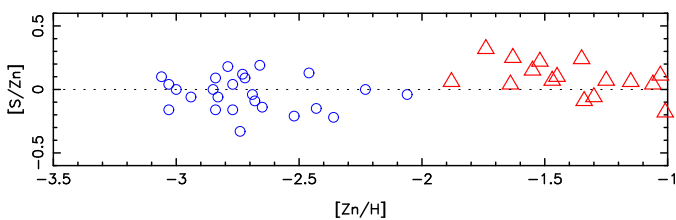}}
\caption{
[S/Zn] vs. [Zn/H] in EMP stars (circles) and in metal-poor DLAs (triangles). }
\label {szn}
\end {center}
\end {figure}

\begin {table}[t] 
\caption {[Zn/H] and [S/Zn] in the most metal-poor DLAs. }  
\label {dlas} 
\begin{center}
\begin{tabular}{l r r r r r }
     Ident   & [Zn/H]&    [S/Zn]&       reference S  &  reference Zn   \\
\hline     
    0010-0012&  -1.35&      0.24&       Srianand,05  &  Ledoux,03      \\ 
    0058-2914&  -1.52&      0.22&       Srianand,05  &  Pettini,00     \\ 
     0100+130&  -1.55&      0.15&       Molaro,98    &  Dessauges-,04  \\ 
    0102-1902&  -1.88&      0.06&       Srianand,05  &  Ledoux,03      \\ 
     0112+029&  -1.01&     -0.18&       Ledoux,03    &  Pettini,94     \\   
     0347-383&  -1.15&      0.06&       Srianand,05  &  Ledoux,03      \\ 
     0405-443&  -1.34&     -0.09&       Lopez,03     &  Ledoux,06      \\ 
     0405-443&  -1.06&      0.04&       Lopez,03     &  Ledoux,06      \\ 
    0528-2505&  -1.47&      0.07&       Centurion,03 &  Ledoux,06      \\  
     0841+129&  -1.45&      0.10&       Ledoux,06    &  Centurion,03   \\  
     0841+129&  -1.74&      0.32&       Centurion,03 &  Dessauges-,06  \\ 
     1223+178&  -1.63&      0.25&       Dessauges-,07&  Ledoux,03      \\   
     1331+170&  -1.25&      0.07&       Dessauges-,04&  Dessauges-,04  \\ 
    2138-4427&  -1.64&      0.04&       Srianand,05  &  Ledoux,03      \\ 
     2314-409&  -1.03&      0.11&       Ellison,01   &  Ledoux,06      \\
    2343+1232&  -1.30&     -0.06&       Noterdaeme,07&  Ledoux,06      \\
\hline   
\end {tabular}  
\end {center}  
\end {table}   
     
Nissen et al.  (\cite{NCA04}) suggested that the [S/Zn] ratio is
higher in stars than in damped $\rm Ly\alpha$ systems.  Neither S nor
Zn shows much affinity for dust, although MgS has been recently
proposed as a possible grain constituent by Zhang et al.
(\cite{ZJL09}).  Thus, depletion onto dust in DLAs probably cannot
explain this difference.  
Nissen et al. (\cite{NAA07}) have shown that in the interval $\rm -2.5<[Zn/H]<0$ 
this effect is considerably reduced 
if one adopts the non-LTE corrections to the abundances of S and Zn.\\
Fig.  \ref{szn} shows [S/Zn] vs.  [Zn/H] for
our sample of metal-poor stars and for the most metal-poor DLAs with
both Zn and S abundance determinations (Table \ref{dlas}).  Since the
Zn line is very weak in turnoff stars, we could only measure Zn
abundances in our giant stars (Cayrel et al., \cite{CDS04}).
Following Takeda et al.  (\cite{THT05}), non-LTE effects for Zn are
not very important in these stars ($\rm \leq +0.1 dex$), but we have
applied  the small correction obtained by interpolation in their tables.
 
With stellar NLTE effects taken into account for both S and Zn,
Fig.  \ref{szn} shows that the sulfur abundance in metal-poor stars is rather similar
to that in DLAs -- if anything slightly lower.  However, the chemical
evolution of the Galactic halo might still be different from the
average of DLAs.

In EMP stars as in DLAs, the [S/Zn] ratio is approximately solar.  This
is due to positive values of both [S/Fe] (Fig.  \ref {S-Fe}) and
[Zn/Fe] at very low metallicity (Cayrel et al.  \cite{CDS04}; see also
Primas et al.  \cite{PBS00} and Nissen et al.  \cite{NCA04}).  We will
discuss the evolution of the [S/Zn] ratio in the Galaxy in more detail
when we have completed a new precise analysis of S abundances in less
metal-poor stars.

\section{Conclusions}
We have determined S abundances in 33 metal-poor stars, including 21
with $\rm[Fe/H] < -2.9$, and find [S/Fe] to be fairly constant at
$\rm[S/Fe]=0.35\pm 0.10$.  However, a slight increase in [S/Mg] when
[Mg/H] decreases cannot be excluded (Fig.  \ref{smg}).  The same
effect is also observed  for Na, Al, and K (Spite et al.  \cite{SSB09},
Andrievsky et al.  \cite{ASK10}).

We do not confirm earlier claims of a significant rise in [S/Fe] with
decreasing [Fe/H], but corroborate the findings of Nissen et al.
(\cite{NCA04}, \cite{NAA07}) and Ryde \& Lambert (\cite{RL04}) that S
behaves like the other $\alpha$-elements, also at very low
metallicity.

The low sulfur abundance (along with a low abundance of zinc) in
CS~22949-037 suggests that the abundance pattern in this star (and
thus by analogy in the hyper iron-poor stars HE~0107-5240 and
HE~1327-2326) is the result of specific zero-metal supernovae 
and is unrelated to dust depletion as suggested by Venn \& Lambert
(\cite{VL08}).

When the NLTE effects are taken into account, the pattern of the light
elements from O to K relative to magnesium, is rather well represented
by ejecta of zero-metal supernovae (Heger \& Woosley \cite{HW10}).

The ratio [S/Zn] is found to be approximately solar as it is in DLAs
since [S/Fe] and [Zn/Fe] are both equally positive at low metallicity.

\begin{acknowledgements} SMA gratefully acknowledges the support and
hospitality of Observatoire de Paris-Meudon.  MS, EC, FS, PB, RC, PF
and VH acknowledge the support of CNRS (PNCG and PNPS).  BB
acknowledges grants from FAPESP and CNPq.  BN and JA acknowledge the
support of the Carlsberg Foundation and the Danish Natural Science
Research Council and TCB acknowledges partial funding of this work
from grants PHY 02-16783 and PHY 08-22648: Physics Frontiers
Center/Joint Institute for Nuclear Astrophysics (JINA), awarded by the
U.S. National Science Foundation. We also thank the anonymous referee 
for his helpful comments.

\end{acknowledgements}

\end{document}